\documentclass[12pt]{article}
\usepackage{epsfig}
\def\PLB{{ Phys. Lett.}  B}
\def\PRL{ Phys. Rev. Lett.}
\def\PRD{{ Phys. Rev.} D}

\def\Journal#1#2#3#4{{#1} {\bf #2}, #3~(#4)}
\newcommand{\mysection}{\setcounter{equation}{0}\section}

\def\beq{\begin{equation}}
\def\eeq{\end{equation}}
\def\beqa{\begin{eqnarray}}
\def\eeqa{\end{eqnarray}}

\newlength{\dinwidth} \newlength{\dinmargin}
\begin{document}
\begin {flushright}
Cavendish-HEP-03/27\\
FSU-HEP-031021\\
\end {flushright} 
\vspace{3mm}
\begin{center}
{\Large \bf FCNC top quark production via anomalous $tqV$ couplings
beyond leading order}
\end{center}
\vspace{2mm}
\begin{center}
{\large Nikolaos Kidonakis$^a$ and Alexander Belyaev$^b$}\\
\vspace{2mm}
$^a${\it Cavendish Laboratory, University of Cambridge,\\
Madingley Road, Cambridge CB3 0HE, UK}\\
\vspace{2.5mm}
$^b${\it Physics Department, Florida State University,\\
    Tallahassee, FL 32306-4350, USA }
\end{center}

\begin{abstract}
We calculate flavor-changing neutral-current (FCNC) processes
involving top-quark production via anomalous $tqV$ couplings
at the Tevatron and HERA colliders.
We cover the FCNC  processes 
$p\bar{p}\to tZ$, $p\bar{p}\to t\gamma$, $p\bar{p}\to t t$, and $ep\to e t$.
We go beyond leading order and include the effects of
soft-gluon corrections through next-to-next-to-leading order.
We demonstrate the stabilisation of the
cross section with respect to the variation of QCD scale,
and we investigate the reach of the Tevatron and HERA colliders.
\end{abstract}

\thispagestyle{empty} 

\newpage \setcounter{page}{2}

\mysection{Introduction}
The top quark, the heaviest known fermion,
occupies a unique place in the search for new physics
beyond the Standard Model (SM).
This is the only fermion with a mass of about the scale
of the electroweak symmetry breaking (EWSB). Therefore the study 
of its properties and their possible deviations from SM
predictions could shed a light on the mechanism of EWSB.
Top quark physics can probe various physics beyond the SM:
anomalous gluon-top quark
couplings~\cite{glu-top},  anomalous $Wtb$ couplings~\cite{sngl-wtb},  
new strong dynamics~\cite{sngl-strong}, 
flavor-changing neutral-currents (FCNC)~\cite{sngl-fcnc}, R-parity
violating SUSY effects~\cite{sngl-rpv}, CP-violation
effects~\cite{sngl-cp}, and 
effects of Kaluza-Klein excited $W$-bosons~\cite{sngl-extra}.

In particular, the study of FCNC couplings involving the top quark
is well motivated.
The effective Lagrangian involving such  couplings of a $t,q$ pair 
to massless bosons is the following:
\begin{equation}
\Delta {\cal L}^{eff} =    \frac{1}{ \Lambda } \,
\kappa_{tqV} \, e \, \bar t \, \sigma_{\mu\nu} \, q \, F^{\mu\nu}_V + h.c.,
\label{fcnc-eq}
\end{equation}
where $\kappa_{tqV}$ is the anomalous FCNC coupling, with 
$q$ a $u$- or $c$-quark and $V$ a photon or $Z$-boson;
$F^{\mu\nu}_V$  are the usual photon/Z-boson field tensors;
$\sigma_{\mu \nu}=(i/2)(\gamma_{\mu}\gamma_{\nu}
-\gamma_{\nu}\gamma_{\mu})$ with $\gamma_{\mu}$ the Dirac matrices;
$e$ is the electron charge; and $\Lambda$ is an effective scale which 
we will take throughout this
paper to be equal to the top quark mass, which we denote by $m$.
In the SM these operators can be induced through higher-order loops, 
however these effects are too small to be observable~\cite{fcnc-sm}.

In some models involving new physics FCNC top-quark couplings 
could appear at tree-level and, therefore, lead to large
FCNC effects. 
Such enchancements of FCNC couplings involving top quarks
could appear in  two-Higgs doublet models and supersymmetric 
models~\cite{fcnc-2hd-mssm} as well as in 
multiple-Higgs doublet models~\cite{fcnc-mhd}.
Models with FCNC coupled singlet quarks, dynamical 
electroweak symmetry breaking, or compositeness, can further
enchance FCNC  top-quark couplings~\cite{fcnc-new-dyn}.
 
The presently operating  colliders such as the Tevatron 
and HERA have a nice opportunity to probe  FCNC 
interactions in the top-quark sector. In order to establish accurate limits
on FCNC couplings, experiments need accurate predictions of cross sections
for FCNC processes. We have shown in~\cite{fcnc-hera} that at HERA 
uncertainties for the tree-level FCNC cross section 
$ep \rightarrow et$ can be very big, as 
the cross section can vary by a factor of two 
due to the choice of the QCD scale. This fact unavoidably requires involving
the higher-order corrections for the stabilization of the FCNC cross section.

In this paper we study some principal FCNC processes both for the Tevatron 
and HERA colliders involving  $t-u-V$ couplings with
$V$ a photon or $Z$-boson \cite{tZgamprod}.  
The effects of $\gamma-Z$ interference
are also taken into account in our studies.
We include soft-gluon corrections at next-to-leading order (NLO)
and next-to-next-to-leading order (NNLO).
We  present the stabilization of the 
NNLO cross sections with respect to the variation
of the renormalization/factorization scales.

In Sections~2 and ~3
we study FCNC processes of the associated $top$-quark 
production with $Z$-boson and photon at the Tevatron, respectively.
Section~4 is devoted  to the $uu\to tt$ process
of same-sign top quark production at the Tevatron.
Section~5 extends our previous study 
on single $top$-quark production at HERA, and we now
consider this process at NNLO level as well as study
both $t-u-\gamma$ and $t-u-Z$ vertexes and 
take into account $\gamma-Z$ interference.
Finally in Section~6 we draw our conclusions.

We note that contributions involving the charm quark
via $t-c-V$ couplings are greatly suppressed because of the 
small charm parton densities. In each section we make a note about 
the contribution from charm quarks. 
We also note that we don't consider $t {\bar t}$ production
and $tq$ or $t{\bar q}$ production in this paper because
the Standard Model cross sections for these processes overwhelm 
any FCNC contributions.

In the present paper we use the notation  $\hat{\sigma}$ for
parton-level differential cross sections, while $\sigma$ denotes
the hadronic cross sections.
We also define a kinematical variable $s_4$, for each subprocess,
that goes to zero at threshold. The effects of the soft-gluon radiation
appear in the form of logarithmic ``plus'' distributions with respect to $s_4$
of the type
\beq
\left[\frac{\ln^{k}(s_4/m^2)}{s_4} \right]_+, \hspace{10mm} k\le 2n-1\, ,
\label{zplus}
\eeq
at $n$-th order in the QCD coupling $\alpha_s$. 
These plus distributions are the remnants of cancellations of 
infrared divergences between
soft and virtual contributions to the cross section.
We define the leading logarithms (LL) with $k=2n-1$ and the 
next-to-leading logarithms (NLL) with $k=2n-2$.

The plus distributions are 
defined by their integral with any smooth function $f$, such as
parton distributions, as 
\beqa
\int_0^{s_{4 \, max}} ds_4 \, f(s_4) \left[\frac{\ln^k(s_4/m^2)}
{s_4}\right]_{+} &=&
\int_0^{s_{4\, max}} ds_4 \frac{\ln^k(s_4/m^2)}{s_4} [f(s_4) - f(0)]
\nonumber \\ &&
{}+\frac{1}{k+1} \ln^{k+1}\left(\frac{s_{4\, max}}{m^2}\right) f(0) \, .
\label{splus}
\eeqa
They provide significant and dominant contributions to the cross section 
near threshold, where there is limited phase space
for real gluon emission,
as has been shown for many Standard Model processes, including
top production \cite{NK}, direct photon production \cite{NKJO}, 
jet production \cite{NKJOjet}, and 
$W$-boson production \cite{NKASV}. 
A unified approach for the calculation of these soft-gluon corrections 
for various processes in hadron-hadron and lepton-hadron colliders 
has recently been presented in Ref. \cite{NKuni}. 
We follow that reference in calculating
soft-gluon corrections to FCNC processes at NLO and NNLO
with NLL accuracy, i.e. keeping LL and NLL.
We work in the $\overline{\rm MS}$ scheme throughout the paper.
  
\mysection{$gu \rightarrow t Z$}

\begin{figure}[htb] 
\centerline{\epsfxsize=0.6\textwidth\epsffile{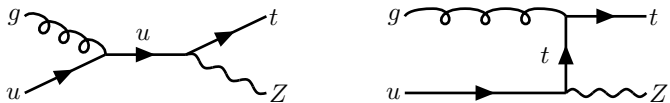}}
\caption[]{Tree-level Feynman diagrams for the process $gu \rightarrow tZ$.}
\label{fig-tz} 
\end{figure}

We begin with FCNC $tZ$ production at the Tevatron.
For the process $g(p_g)+u(p_u) \rightarrow t(p_t)+Z(p_Z)$,
we define the kinematical invariants $s=(p_g+p_u)^2$,
$t=(p_g-p_t)^2$, $u=(p_u-p_t)^2$, and $s_4=s+t+u-m^2-m_Z^2$,
where $m_Z$ is the $Z$-boson mass and $m$ is the top quark mass. 
Note that near threshold, i.e. when we have just enough
partonic energy to produce the $tZ$ final state,  $s_4 \rightarrow 0$.
The respective tree-level Feynman diagrams are shown in Fig.~\ref{fig-tz}.

The differential Born cross section is 
\beq
{\label{eq1}
\frac{d^2{\hat\sigma}^{gu \rightarrow t Z}_B}{dt \; du}
=F^{gu \rightarrow t Z}_B \delta(s_4)
}
\eeq
where 
\beqa
F^{gu \rightarrow t Z}_B&=&
\frac{2 \pi \alpha \alpha_s \kappa_Z^2}{3m^2s^3(m^2-t)^2}
\left\{2 m^8-m^6(3 m_Z^2+4 s+2 t) \right. 
\nonumber \\ &&
{}-t\left[2 m_Z^6-2m_Z^4(s+t)-4 s t(s+t)
+m_Z^2(s+t)^2\right]
\nonumber \\ &&
{}+m^4\left[2m_Z^4-m_Z^2(2s+t)
+2(s^2+4st+t^2)\right] 
\nonumber \\ && \left. 
{}+m^2\left[2m_Z^6-4m_Z^4t+m_Z^2(s+t)(s+5t)
-2t(3s^2+6st+t^2)\right]\right\} \, ,
\eeqa
where $\alpha=e^2/(4\pi)$ and for brevity we define 
$\kappa_Z \equiv \kappa_{tuZ}$.

The NLO soft-gluon corrections for $g u \rightarrow tZ$ are
\beq
\frac{d^2{\hat\sigma}^{(1)}_{gu\rightarrow t Z}}{dt \, du}
=F^{gu \rightarrow t Z}_B 
\frac{\alpha_s(\mu_R^2)}{\pi} \left\{
c^{gu \rightarrow tZ}_{3} \left[\frac{\ln(s_4/m^2)}{s_4}\right]_+
+c^{gu \rightarrow tZ}_{2} \left[\frac{1}{s_4}\right]_+
+c^{gu \rightarrow tZ}_{1}  \delta(s_4)\right\} \, .
\label{NLOgutZ}
\eeq

Here $c^{gu \rightarrow tZ}_{3}=2(C_F+C_A)$, where $C_F=(N_c^2-1)/(2N_c)$
and $C_A=N_c$ with $N_c=3$ the number of colors,  and
\beqa
c^{gu \rightarrow tZ}_{2}&=&2 {\rm Re} {\Gamma'}_S^{(1)}
-C_F-C_A-2C_F\ln\left(\frac{-t+m_Z^2}{m^2}\right)
-2C_A\ln\left(\frac{-u+m_Z^2}{m^2}\right)
-(C_F+C_A)\ln\left(\frac{\mu_F^2}{s}\right)
\nonumber \\ 
& \equiv & T^{gu \rightarrow tZ}_{2}-(C_F+C_A)
\ln\left(\frac{\mu_F^2}{m^2}\right) \, ,
\eeqa
where $\mu_F$ is the factorization scale, and we have defined 
$T^{gu \rightarrow tZ}_{2}$ as the scale-independent part
of $c^{gu \rightarrow tZ}_{2}$.
The term ${\rm Re} {\Gamma'}_S^{(1)}$ denotes the real part
of the one-loop soft anomalous dimension, which describes
noncollinear soft-gluon emission \cite{KOS}. A one-loop 
calculation gives 
\beq
{\Gamma'}_S^{(1)}=C_F \ln\left(\frac{-u+m^2}{m\sqrt{s}}\right)
+\frac{C_A}{2} \ln\left(\frac{-t+m^2}{-u+m^2}\right)
+\frac{C_A}{2} (1-\pi i)\, .
\label{Gamma}
\eeq

Also
\beq
c^{gu \rightarrow tZ}_{1}=\left[
C_F \ln\left(\frac{-t+m_Z^2}{m^2}\right)+C_A \ln\left(\frac{-u+m_Z^2}{m^2}\right)
-\frac{3}{4}C_F-\frac{\beta_0}{4}\right]
\ln\left(\frac{\mu_F^2}{m^2}\right)
+\frac{\beta_0}{4} \ln\left(\frac{\mu_R^2}{m^2}\right)  \, ,
\eeq
where $\mu_R$ is the renormalization scale and $\beta_0=(11C_A-2n_f)/3$ 
is the lowest-order $\beta$ function, 
with $n_f=5$ the number of light quark flavors.
Note that $c^{gu \rightarrow tZ}_{1}$ represents the scale-dependent
part of the $\delta(s_4)$ corrections. We do not calculate the full virtual
corrections here. Our calculation of the soft-gluon corrections
includes the leading and next-to-leading logarithms (NLL) and is thus
a NLO-NLL calculation. 

We next calculate the NNLO soft-gluon corrections for $g u \rightarrow tZ$:
\beqa
&& \hspace{-5mm}\frac{d^2{\hat\sigma}^{(2)}_{gu \rightarrow tZ}}
{dt \, du}
=F^{gu \rightarrow tZ}_B \frac{\alpha_s^2(\mu_R^2)}{\pi^2} 
\left\{\frac{1}{2} \left(c^{gu \rightarrow tZ}_{3}\right)^2 
\left[\frac{\ln^3(s_4/m^2)}{s_4}\right]_+ \right.
\nonumber \\ && \hspace{-5mm}
{}+\left[\frac{3}{2} c^{gu \rightarrow tZ}_{3} \, c^{gu 
\rightarrow tZ}_{2}
-\frac{\beta_0}{4} c^{gu \rightarrow tZ}_{3} \right] 
\left[\frac{\ln^2(s_4/m^2)}{s_4}\right]_+
\nonumber \\ && \hspace{-5mm}
{}+\left[c^{gu \rightarrow tZ}_{3} \, c^{gu \rightarrow tZ}_{1}
+(C_F+C_A)^2\ln^2\left(\frac{\mu_F^2}{m^2}\right)
-2(C_F+C_A) T_2^{gu \rightarrow tZ}\ln\left(\frac{\mu_F^2}{m^2}\right)
\right.
\nonumber \\ && \quad \left.
{}+\frac{\beta_0}{4} c^{gu \rightarrow tZ}_{3} 
\ln\left(\frac{\mu_R^2}{m^2}\right)
-\zeta_2 \, \left(c^{gu \rightarrow tZ}_{3}\right)^2 \right]
\left[\frac{\ln(s_4/m^2)}{s_4}\right]_+
\nonumber \\ && \hspace{-5mm} 
{}+\left[-(C_F+C_A) \ln\left(\frac{\mu_F^2}{m^2}\right)
c^{gu \rightarrow tZ}_{1}
-\frac{\beta_0}{4} (C_F+C_A) \ln\left(\frac{\mu_F^2}{m^2}\right) 
\ln\left(\frac{\mu_R^2}{m^2}\right) \right.
\nonumber \\ && \quad \left. \left.
{}+(C_F+C_A)\frac{\beta_0}{8} \ln^2\left(\frac{\mu_F^2}{m^2}\right)
-\zeta_2 \, c^{gu \rightarrow tZ}_{2} \, c^{gu \rightarrow tZ}_{3}
+\zeta_3 \, \left(c^{gu \rightarrow tZ}_{3}\right)^2\right]
\left[\frac{1}{s_4}\right]_+ \right\} \, ,
\label{NNLOgutZ}
\eeqa
where $\zeta_2=\pi^2/6$ and $\zeta_3=1.2020569...$.
We note that only the leading and next-to-leading logarithms are complete.
Hence this is a NNLO-NLL calculation.
Consistent with a NLL calculation we have also kept all logarithms of the 
factorization and renormalization scales in the
$[\ln(s_4/m^2)/s_4]_+$ terms, and squares of scale logarithms
in the $[1/s_4]_+$ terms, as well as $\zeta_2$ and $\zeta_3$ terms
that arise in the calculation of the soft corrections.
For relevant details in the related process of Standard Model top
production see Ref. \cite{NK}. 

We now convolute the partonic cross sections with parton distribution
functions (PDF) to obtain the hadronic cross section.
For the FCNC hadronic cross section $p(p_a)+{\bar p}(p_b) 
\rightarrow t(p_t)+Z(p_Z)$ we define 
$S=(p_a+p_b)^2$, $T=(p_a-p_t)^2$, and $U=(p_b-p_t)^2$,
and note that $p_g=x_a p_a$, $p_u=x_b p_b$, where $x$ denotes
the momentum fraction of the hadron carried by the parton.
The hadronic cross section is then given by
\beqa
\sigma^{FCNC}_{p{\bar p} \rightarrow tZ}(S)&=&
\int_{T_{min}}^{T_{max}} dT 
\int_{-S-T+m^2+m_Z^2}^{m^2+m^2S/(T-m^2)} dU 
\int_{(m_Z^2-T)/(S+U-m^2)}^1 dx_b \int_0^{x_b(S+U-m^2)+T-m_Z^2} ds_4
\nonumber \\ &&
\times \frac{x_a x_b}{x_b S+T-m^2} \,
\phi(x_a) \, \phi(x_b) \, 
\frac{d^2{\hat\sigma}_{gu \rightarrow tZ}}{dt \, du}
\eeqa
where
\beq
T_{^{max}_{min}}=-\frac{1}{2}(S-m^2-m_Z^2) \pm 
\frac{1}{2} \sqrt{(S+m^2-m_Z^2)^2-4m^2S} \, ,
\eeq
$x_a=\left[s_4-m^2+m_Z^2-x_b(U-m^2)\right]/(x_b S+T-m^2)$, and
$\phi(x)$ are the parton distributions.

In our calculations we are using the MRST2002 NNLO
parton distribution functions ~\cite{mrst2002}
with the respective  three-loop evaluation of $\alpha_s$. 
For all three $-$  Born, NLO-NLL,  and NNLO-NLL $-$
we use the same  MRST2002 NNLO PDF.
We also use $\mu_F=\mu_R$ for our numerical results.
For reference point we choose $\kappa_Z$=0.1 and
$\mu \equiv \mu_F=\mu_R=m=175$~GeV.
In our paper we quote the theoretical error related
to the conventional variation of the scale between $m/2$
and $2m$.

\begin{figure}[htb] 
\epsfxsize=0.5\textwidth\epsffile{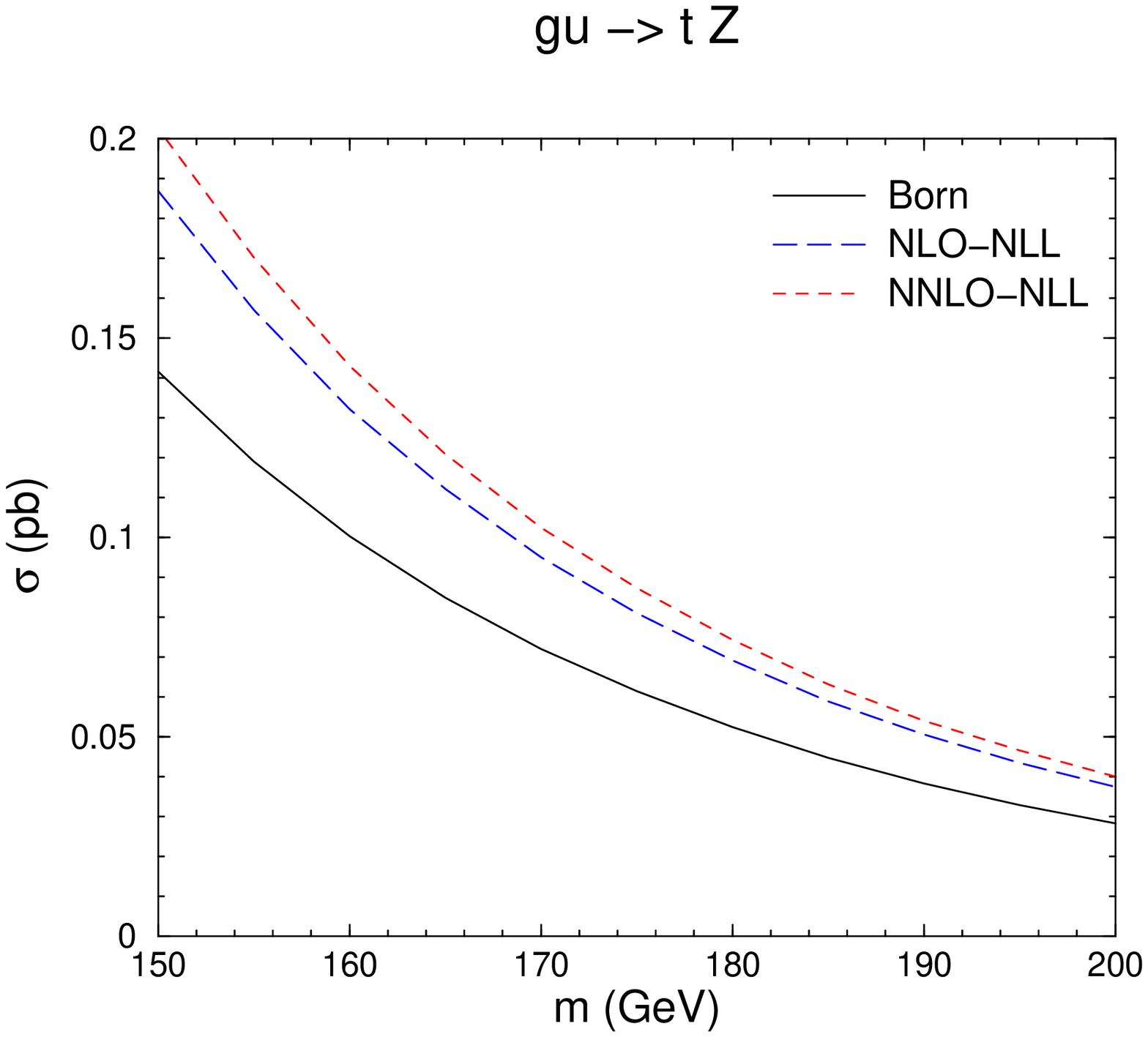}
\epsfxsize=0.5\textwidth\epsffile{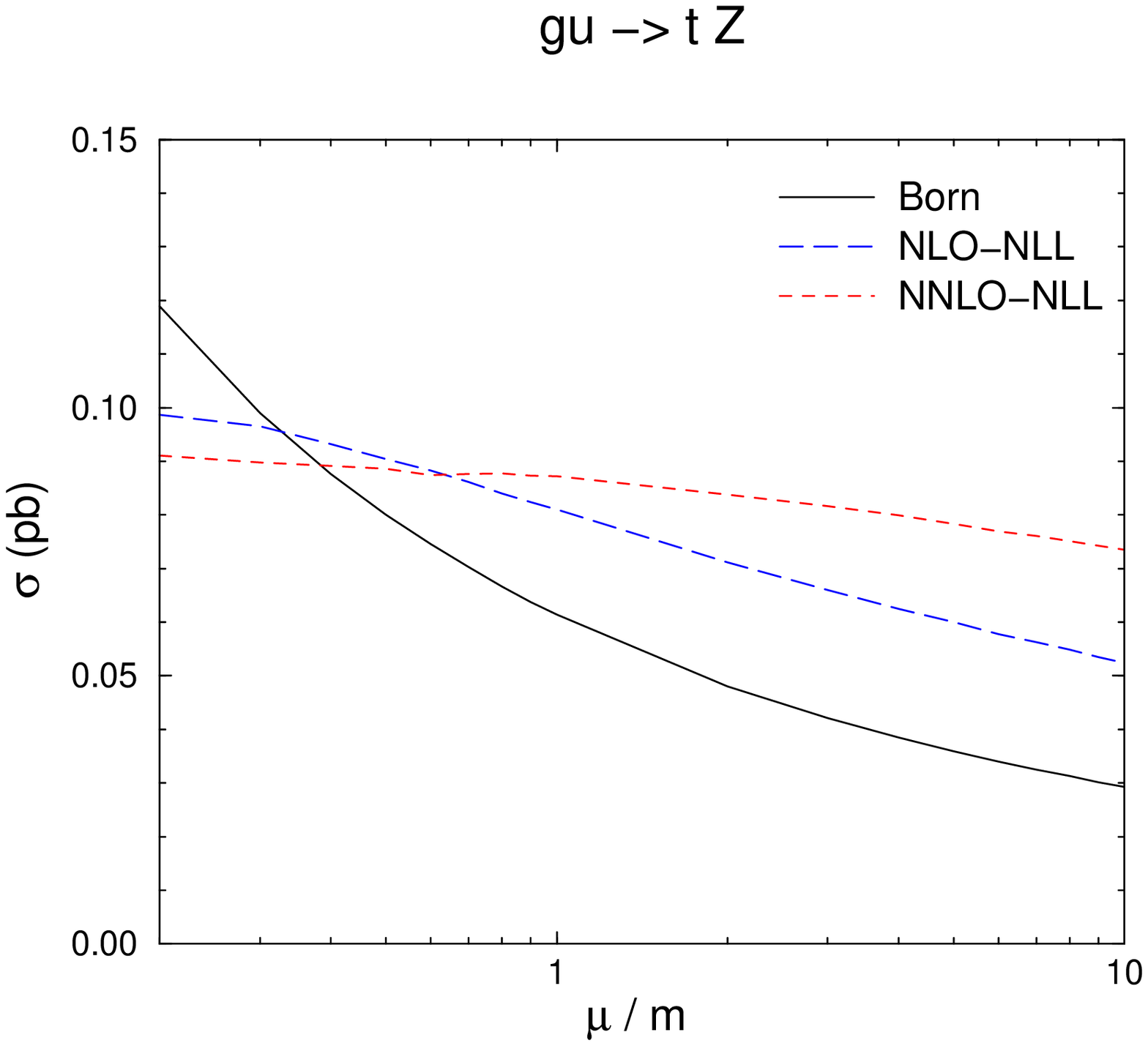}
\caption[]{The Born, NLO-NLL, and NNLO-NLL cross sections
for $gu\rightarrow tZ$ in $p \bar p$ collisions
with $\sqrt{S}=1.96$ TeV and $\kappa_Z=0.1$.
Left:  cross section versus top-quark mass at scale $\mu=m$;
right: cross section versus  the   scale for $m= 175$~GeV.
}
\label{fcnc-cs-tz} 
\end{figure}

We present our numerical results  in Fig.~\ref{fcnc-cs-tz}.
The left frame presents  the cross section for the process  
$gu\rightarrow t Z$ versus top-quark mass at the Tevatron
with $\sqrt{S}=1.96$ TeV, at Born, NLO-NLL, and NNLO-NLL
orders. Note that the first-order corrections, $\Delta\sigma_{NLO-NLL}$, 
and the second-order corrections, $\Delta\sigma_{NNLO-NLL}$, 
are positive for $\mu=m$.
There is clear stabilisation of the  cross section 
due to the $\Delta\sigma_{NLO-NLL}$ and $\Delta\sigma_{NNLO-NLL}$
corrections over a large range of $\mu/m$ presented  in the
right frame of Fig.~\ref{fcnc-cs-tz}. 
At the reference point, $\mu=m=175$~GeV and $\kappa_Z=0.1$,
we have the following results:
$$\sigma_{Born}^{gu\rightarrow t Z}          = 61\ {\rm fb}, 
\ \Delta\sigma_{NLO-NLL}^{gu\rightarrow t Z}= 20\ {\rm fb}, 
\ \Delta\sigma_{NNLO-NLL }^{gu\rightarrow t Z}= 6 \ {\rm fb}.$$
Taking into account the theoretical error due to the scale variation
between $m/2$ and $2m$ one thus has for the total NNLO-NLL cross section:
$$\sigma_{NNLO-NLL }^{gu\rightarrow t Z} =87^{+2}_{-3}\ {\rm fb}.$$

Since the total cross section is proportional to $\kappa_Z^2$
it is straightforward to rescale it for any given value of 
the anomalous FCNC coupling.

For completeness we would like also to quote 
the cross section for the case of  $\kappa_{tcZ}$
coupling. Since the initial $c$-quark is coming from the sea, 
and thus the charm density is quite small,
the total cross section for $gc\rightarrow t Z$ at the Tevatron, 
with $\mu=m=175$ GeV and assuming the anomalous coupling is again $0.1$, 
is about 40 times smaller:
$\sigma_{NNLO-NLL }^{gc\rightarrow t Z} 
=2.4\ {\rm fb}$.

One should also notice that 
we present all cross sections for $top$-quark production.
For experimental studies one should also include
the contribution from  $anti$-$top$-quark
production which will double the total
cross section for all processes at the Tevatron we study here.

\mysection{$gu \rightarrow t \gamma$}

\begin{figure}[htb] 
\centerline{\epsfxsize=0.6\textwidth\epsffile{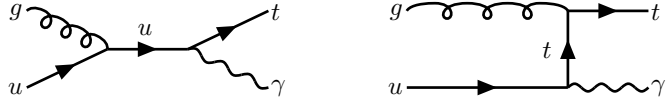}}
\caption[]{Tree-level Feynman diagrams for the process 
$gu \rightarrow t \gamma$.}
\label{fig-tgam} 
\end{figure}

We continue with FCNC $t \gamma$ production at the Tevatron.
For the process $g(p_g)+u(p_u) \rightarrow t(p_t)+\gamma(p_{\gamma})$,
we define the kinematical invariants $s=(p_g+p_u)^2$,
$t=(p_g-p_t)^2$, $u=(p_u-p_t)^2$, and $s_4=s+t+u-m^2$, 
where again $m$ is the top-quark mass.
Note that near threshold $s_4 \rightarrow 0$.
The respective tree-level Feynman diagrams are shown in Fig.~\ref{fig-tgam}.

The differential Born cross section is 
\beq
\frac{d^2{\hat\sigma}^{gu \rightarrow t \gamma}_B}{dt \; du}
=F^{gu \rightarrow t \gamma}_B \delta(s_4)
\eeq
where 
\beq
F^{gu \rightarrow t \gamma}_B=
\frac{4 \pi \alpha \alpha_s  \kappa_{\gamma}^2(m^2-s-t)
\left[m^6-m^4 s-2 s t^2+m^2 t (3 s+t)\right]}
{3 \, m^2 \, s^3 \, (m^2-t)^2} \, ,
\eeq
where we define $\kappa_{\gamma} \equiv \kappa_{tq\gamma}$.

The NLO-NLL corrections for $g u \rightarrow t \gamma$ are
\beq
\frac{d^2{\hat\sigma}^{(1)}_{gu\rightarrow t \gamma}}{dt \, du}
=F^{gu \rightarrow t \gamma}_B 
\frac{\alpha_s(\mu_R^2)}{\pi} \left\{
c^{gu \rightarrow t\gamma}_{3} \left[\frac{\ln(s_4/m^2)}{s_4}\right]_+
+c^{gu \rightarrow t\gamma}_{2} \left[\frac{1}{s_4}\right]_+
+c^{gu \rightarrow t\gamma}_{1}  \delta(s_4)\right\} \, .
\label{NLOgutgamma}
\eeq

Here $c^{gu \rightarrow t\gamma}_{3}=2(C_F+C_A)$, 
\beqa
c^{gu \rightarrow t\gamma}_{2}&=&2 {\rm Re} {\Gamma'}_S^{(1)}
-C_F-C_A-2C_F\ln\left(\frac{-t}{m^2}\right)
-2C_A\ln\left(\frac{-u}{m^2}\right)
-(C_F+C_A)\ln\left(\frac{\mu_F^2}{s}\right)
\nonumber \\ 
& \equiv & T^{gu \rightarrow t\gamma}_{2}-(C_F+C_A)
\ln\left(\frac{\mu_F^2}{m^2}\right) \, ,
\eeqa
with ${\Gamma'}_S^{(1)}$ defined in Eq. (\ref{Gamma}),
and
\beq
c^{gu \rightarrow t\gamma}_{1}=\left[
C_F \ln\left(\frac{-t}{m^2}\right)+C_A \ln\left(\frac{-u}{m^2}\right)
-\frac{3}{4}C_F-\frac{\beta_0}{4}\right]
\ln\left(\frac{\mu_F^2}{m^2}\right)
+\frac{\beta_0}{4} \ln\left(\frac{\mu_R^2}{m^2}\right)  \, .
\eeq
We note that the $c_i$ coefficients are similar to those for the
$gu \rightarrow tZ$ process. This is because the QCD content
of the two processes is the same.

The NNLO-NLL corrections for $g u \rightarrow t \gamma$ are
\beqa
&& \hspace{-5mm}\frac{d^2{\hat\sigma}^{(2)}_{gu \rightarrow t\gamma}}
{dt \, du}
=F^{gu \rightarrow t\gamma}_B \frac{\alpha_s^2(\mu_R^2)}{\pi^2} 
\left\{\frac{1}{2} \left(c^{gu \rightarrow t\gamma}_{3}\right)^2 
\left[\frac{\ln^3(s_4/m^2)}{s_4}\right]_+ \right.
\nonumber \\ && \hspace{-5mm}
{}+\left[\frac{3}{2} c^{gu \rightarrow t\gamma}_{3} \, c^{gu 
\rightarrow t\gamma}_{2}
-\frac{\beta_0}{4} c^{gu \rightarrow t\gamma}_{3} \right] 
\left[\frac{\ln^2(s_4/m^2)}{s_4}\right]_+
\nonumber \\ && \hspace{-5mm}
{}+\left[c^{gu \rightarrow t\gamma}_{3} \, c^{gu \rightarrow t\gamma}_{1}
+(C_F+C_A)^2\ln^2\left(\frac{\mu_F^2}{m^2}\right)
-2(C_F+C_A) T_2^{gu \rightarrow t \gamma}\ln\left(\frac{\mu_F^2}{m^2}\right)
\right.
\nonumber \\ && \quad \left.
{}+\frac{\beta_0}{4} c^{gu \rightarrow t\gamma}_{3} 
\ln\left(\frac{\mu_R^2}{m^2}\right)
-\zeta_2 \, \left(c^{gu \rightarrow t\gamma}_{3}\right)^2 \right]
\left[\frac{\ln(s_4/m^2)}{s_4}\right]_+
\nonumber \\ && \hspace{-5mm} 
{}+\left[-(C_F+C_A) \ln\left(\frac{\mu_F^2}{m^2}\right)
c^{gu \rightarrow t\gamma}_{1}
-\frac{\beta_0}{4} (C_F+C_A) \ln\left(\frac{\mu_F^2}{m^2}\right) 
\ln\left(\frac{\mu_R^2}{m^2}\right) \right.
\nonumber \\ && \quad \left. \left.
{}+(C_F+C_A)\frac{\beta_0}{8} \ln^2\left(\frac{\mu_F^2}{m^2}\right)
-\zeta_2 \, c^{gu \rightarrow t\gamma}_{2} \, c^{gu \rightarrow t\gamma}_{3}
+\zeta_3 \, \left(c^{gu \rightarrow t\gamma}_{3}\right)^2\right]
\left[\frac{1}{s_4}\right]_+ \right\} \, .
\label{NNLOgutgamma}
\eeqa
Again, they are of the same form as for the $gu \rightarrow tZ$ process. 

\begin{figure}[htb] 
\epsfxsize=0.5\textwidth\epsffile{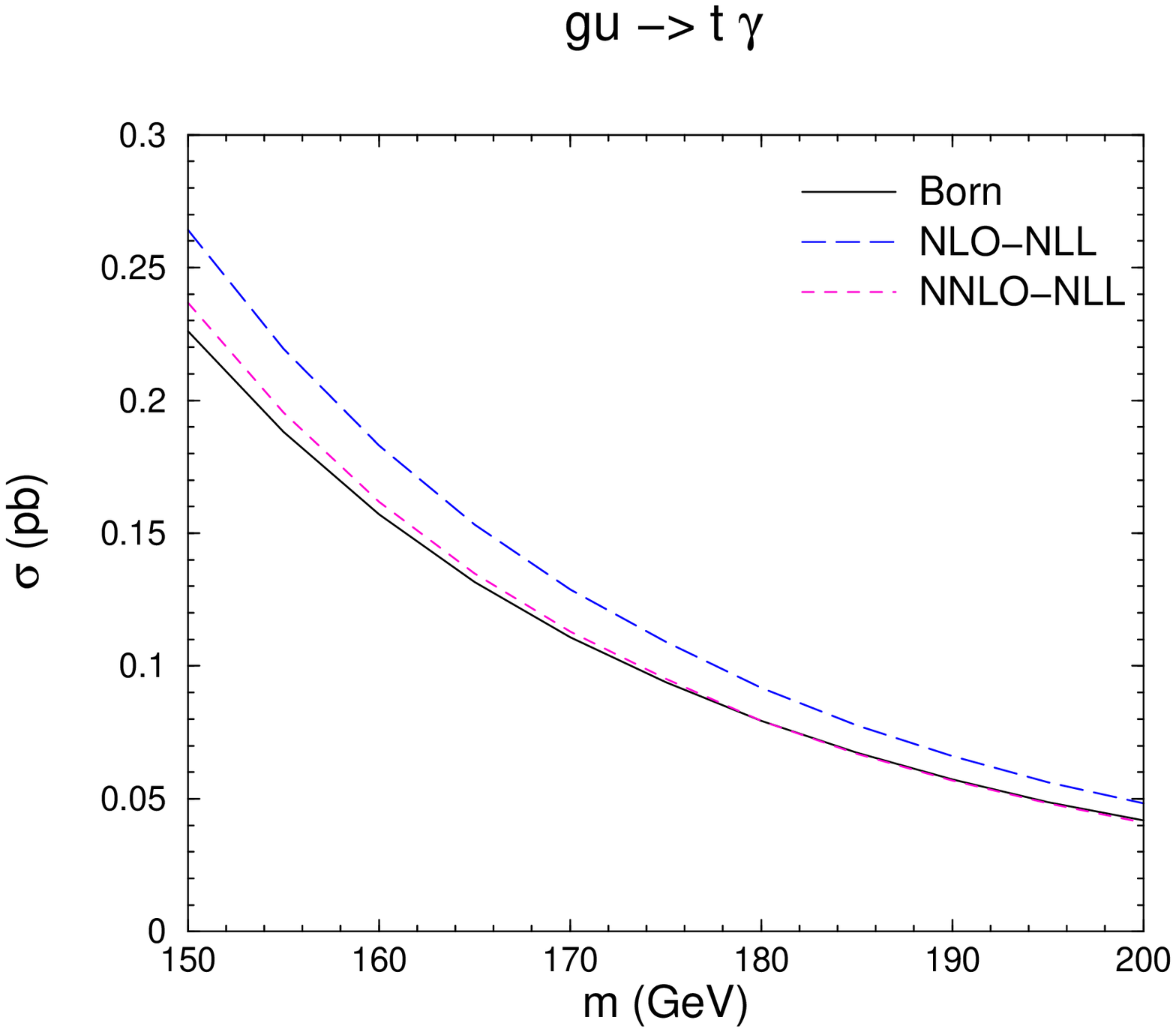}
\epsfxsize=0.5\textwidth\epsffile{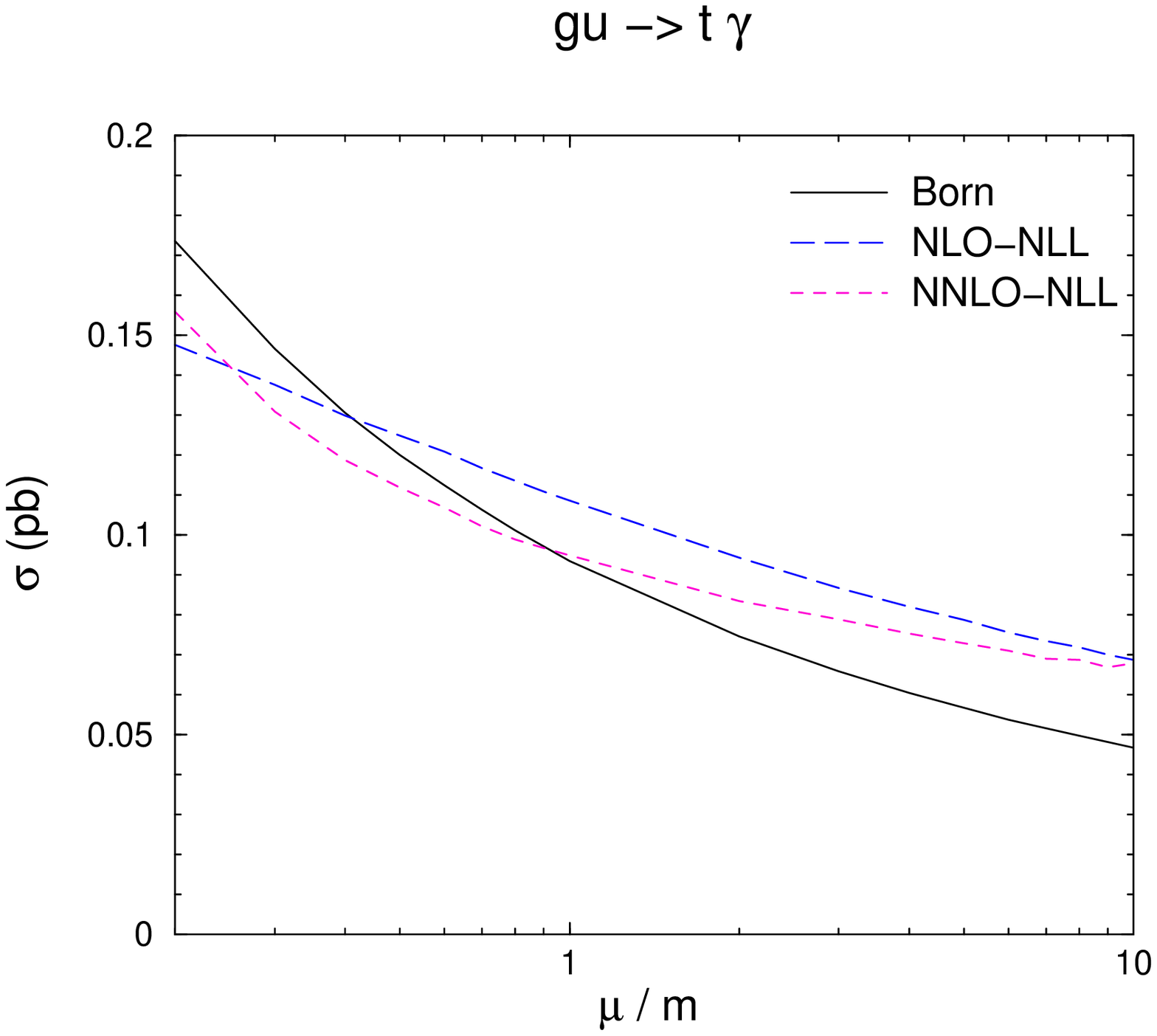}
\caption[]{The Born, NLO-NLL, and NNLO-NLL cross sections 
for $gu\rightarrow t \gamma$ in $p \bar p$ collisions
with $\sqrt{S}=1.96$ TeV and $\kappa_{\gamma}=0.1$.
Left:  cross section versus top-quark mass at scale $\mu=m$;
right: cross section versus  the  scale for $m= 175$~GeV.
}
\label{fcnc-cs-tgam} 
\end{figure}

For the FCNC hadronic cross section $p(p_a)+{\bar p}(p_b) 
\rightarrow t(p_t)+\gamma(p_{\gamma})$
we define $S=(p_a+p_b)^2$, $T=(p_a-p_t)^2$, and $U=(p_b-p_t)^2$,
and note that $p_g=x_a p_a$, $p_u=x_b p_b$.
The hadronic cross section is then given by
\beqa
\sigma^{FCNC}_{p{\bar p} \rightarrow t \gamma}(S)&=&\int_{m^2-S}^0 dT 
\int_{-S-T+m^2}^{m^2+m^2S/(T-m^2)} dU 
\int_{-T/(S+U-m^2)}^1 dx_b \int_0^{x_b(S+U-m^2)+T} ds_4
\nonumber \\ &&
\times \frac{x_a x_b}{x_b S+T-m^2} \, 
\phi(x_a) \, \phi(x_b) \, 
\frac{d^2{\hat\sigma}_{gu \rightarrow t\gamma}}{dt \, du}
\eeqa
with
$x_a=\left[s_4-m^2-x_b(U-m^2)\right]/(x_b S+T-m^2)$.

Our numerical results are presented in Fig.~\ref{fcnc-cs-tgam}.
The left frame presents  the cross section for the process 
$gu\rightarrow t \gamma$ versus top-quark mass at the Tevatron,
with $\sqrt{S}=1.96$ TeV,  at Born, NLO-NLL, and NNLO-NLL
orders. One can see that the NLO-NLL correction is positive at $\mu=m$ 
but the NNLO-NLL correction is negative. 
The right frame presents the scale dependence of the
cross section which again is stabilized when corrections are added.
At $\mu=m=175$~GeV, the Born and  NNLO-NLL results happen to be
very close to each other because of the negative NNLO-NLL correction.
For our reference point, $\mu=m=175$ GeV, $\kappa_{\gamma}=0.1$, we have:
$$\sigma_{Born}^{gu\rightarrow t \gamma}          = 94\ {\rm fb}, 
\ \Delta\sigma_{NLO-NLL}^{gu\rightarrow t \gamma}= 15\ {\rm fb}, 
\ \Delta\sigma_{NNLO-NLL}^{gu\rightarrow t \gamma}=-14\ {\rm fb}.$$
This gives a NNLO-NLL total cross section
$$\sigma_{NNLO-NLL }^{gu\rightarrow t \gamma} 
=95^{+17}_{-11}\ {\rm fb}$$
with theoretical error due to the scale variation
between $m/2$ and $2m$.

Since the total cross section is proportional to $\kappa_{\gamma}^2$
it is straightforward to rescale it for any given value of 
the anomalous FCNC coupling.

For completeness we would like also to quote 
the cross section for the case of  $\kappa_{tc\gamma}$
coupling. Since the initial $c$-quark is coming from the sea,
the total cross section at the Tevatron 
for the process $gc\rightarrow t \gamma$, 
at $\mu=m=175$ GeV and assuming the coupling is again $0.1$,
is about 35 times lower then the one for $gu\rightarrow t \gamma$:
$\sigma_{NNLO-NLL }^{gc\rightarrow t \gamma} 
=2.7\ {\rm fb}$.

Again we  note that we present all cross sections for $top$-quark production.
For experimental studies one should also include
the contribution from  $anti$-$top$-quark
production which will double the total
cross section at the Tevatron.

\mysection{$uu \rightarrow tt$}

\begin{figure}[htb] 
\centerline{\epsfxsize=0.6\textwidth\epsffile{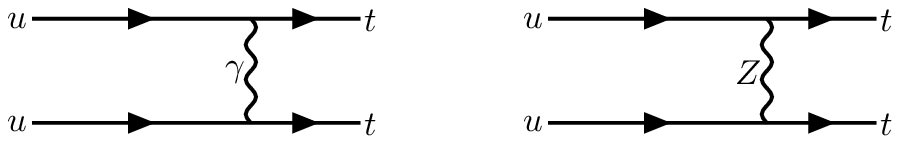}}
\caption[]{Tree-level Feynman diagrams for the process $uu \rightarrow t t$.}
\label{fig-tt} 
\end{figure}

We now turn to FCNC same-sign top-quark production at the Tevatron.
For the process, $u(p_{u_a})+u(p_{u_b}) \rightarrow t(p_1) +t (p_2)$,
we define the kinematical invariants $s=(p_{u_a}+p_{u_b})^2$,
$t=(p_{u_a}-p_1)^2$, $t_1=t-m^2$, $u=(p_{u_b}-p_1)^2$, $u_1=u-m^2$, and 
$s_4=s+t_1+u_1$, with $m$ the top-quark mass. At threshold $s_4 \rightarrow 0$.
The respective tree-level Feynman diagrams are shown in Fig.~\ref{fig-tt}.

The color structure of the hard scattering is somewhat complex since 
we have two top quarks in the final state and two up quarks in 
the initial state. This is in contrast to all the other processes
discussed in this paper which have simple color structure.
We thus decompose the tree-level amplitudes in terms of color tensors
using the methods and techniques detailed in Ref. \cite{KOS}.
As we will see the soft anomalous dimension is now a matrix in color
space.

For the process at hand we choose the color basis consisting of
singlet exchange in the $t$ and $u$ channels,
$c_1=\delta_{a1} \delta_{b2}$ and $c_2=\delta_{a2} \delta_{b1}$,
where $a$ ($b$) is the color index for the $u$-quark with momentum $p_{u_a}$
($p_{u_b}$) and 1 (2) is the color index for the top quark with
momentum $p_1$ ($p_2$). Of course the physical cross section is
independent of the specific choice of color basis.
We write the Born cross section as a trace of
the product of a ``hard'' function, $H$, that describes the short-distance
hard-scattering,  and a ``soft'' function, $S$, that describes
noncollinear soft gluon emission \cite{NK,NKuni,KOS}. 
Both $H$ and $S$ are $2 \times 2$ matrices 
in the color basis $c_1$, $c_2$.

The hard matrix at lowest order, calculated by projecting
the tree-level amplitude onto the color basis, is given by
\beq
H^{(0)}=\frac{1}{16 \pi s^2}\left[\begin{array}{cc}
H_{11}^{(0)} & H_{12}^{(0)}   \vspace{2mm} \\
H_{21}^{(0)} & H_{22}^{(0)}   \vspace{2mm}
\end{array}\right]
\eeq
with
\beqa
H_{11}^{(0)}&=&\frac{8}{9} \frac{(4\pi\alpha)^2}{m^4}
\left[2m^8+4m^6t+t^2(t+2u)^2-m^4t(t+4u)-2m^2t^2(t+4u)\right]
\nonumber \\ && \times 
\left[\frac{\kappa_{\gamma}^4}{t^2}
-\frac{2\kappa_{\gamma}^2 \kappa_Z^2}{t (m_Z^2-t)}
+\frac{\kappa_Z^4}{(m_Z^2-t)^2}\right]
\eeqa
and
\beq
H_{22}^{(0)}=H_{11}^{(0)}(t \leftrightarrow u) \, .
\eeq
Here $\kappa_{\gamma} \equiv \kappa_{tu\gamma}$ and
$\kappa_{Z} \equiv \kappa_{tuZ}$.
Also 
\beqa
H_{12}^{(0)}&=&-\frac{4}{9} \frac{(4\pi\alpha)^2}{m^4} 
\left[8m^8-5m^4tu-2m^6(t+u)+5m^2tu(t+u)-tu(2t+u)(t+2u)\right]
\nonumber \\ && \times 
\left[\frac{\kappa_{\gamma}^4}{tu}
-\frac{2\kappa_{\gamma}^2 \kappa_Z^2}{t (m_Z^2-u)}
+\frac{\kappa_Z^4}{(m_Z^2-t) (m_Z^2-u)}\right]
\eeqa
and 
\beq
H_{21}^{(0)}=H_{12}^{(0)}(t \leftrightarrow u) \, .
\eeq
The soft matrix at lowest order is given by
\beq
S^{(0)}=\left[\begin{array}{cc}
N_c^2 & N_c   \vspace{2mm} \\
N_c   & N_c^2 \vspace{2mm}
\end{array}\right] \, .
\eeq
Note that the matrix elements of $S^{(0)}$ are simply
given by $S^{(0)}_{ji}={\rm Tr}[c_j^* c_i]$.

Then the differential Born cross section is
\beq
\frac{d^2{\hat\sigma}^{uu \rightarrow tt}_B}{dt \; du}
=F^{uu \rightarrow tt}_B \delta(s_4)
\eeq
with
\beq
F^{uu \rightarrow tt}_B={\rm Tr} [H^{(0)} S^{(0)}] \, .
\eeq

The NLO-NLL corrections for $u u \rightarrow tt$ are
\beqa
\frac{d^2{\hat\sigma}^{(1)}_{uu\rightarrow tt}}{dt \, du}
&=&F^{uu \rightarrow tt}_B 
\frac{\alpha_s(\mu_R^2)}{\pi} \left\{
c^{uu \rightarrow tt}_{3} \left[\frac{\ln(s_4/m^2)}{s_4}\right]_+
+c^{uu \rightarrow tt}_{2} \left[\frac{1}{s_4}\right]_+
+c^{uu \rightarrow tt}_{1}  \delta(s_4)\right\}
\nonumber \\ &&
{}+\frac{\alpha_s(\mu_R^2)}{\pi} A^{uu \rightarrow tt} 
\left[\frac{1}{s_4}\right]_+ \, .
\label{NLOuutt}
\eeqa

Here $c^{uu \rightarrow tt}_{3}=4C_F$,
\beq
c^{uu \rightarrow tt}_{2}=-2C_F-2C_F\ln\left(\frac{t_1u_1}{m^4}\right)
-2C_F\ln\left(\frac{\mu_F^2}{s}\right)
\equiv T^{uu \rightarrow tt}_{2}-2C_F\ln\left(\frac{\mu_F^2}{m^2}\right) \, ,
\eeq
where $T^{uu \rightarrow tt}_{2}$ denotes the scale-independent part of
$c^{uu \rightarrow tt}_{2}$, 
and
\beq
c^{uu \rightarrow tt}_{1}=
C_F \left[\ln\left(\frac{t_1u_1}{m^4}\right)-\frac{3}{2}\right]
\ln\left(\frac{\mu_F^2}{m^2}\right) 
\eeq
is again the scale-dependent part of the $\delta(s_4)$ terms.
Also
\beq 
A^{uu \rightarrow tt}={\rm Tr} \left[H^{(0)} {\Gamma'}_S^{(1)\, \dagger} 
S^{(0)}+H^{(0)} S^{(0)} {\Gamma'}_S^{(1)}\right] 
\label{Auutt}
\eeq
is the part of the NLO-NLL corrections not proportional to the Born
cross section; this is again because of the complex color structure
of the hard scattering.
Here ${\Gamma'}_S^{(1)}$ is a $2 \times 2$ soft anomalous dimension matrix,
calculated at one loop, with elements
\beqa
{\Gamma'}_{11}^{(1)}&=&C_F\left[2\ln\left(\frac{-t_1}{s}\right)
-\ln\left(\frac{m^2}{s}\right)\right]-\frac{1}{2N_c}\left[2\ln
\left(\frac{-u_1}{s}\right)-\ln\left(\frac{m^2}{s}\right)
+L_{\beta}+\pi i\right] \, ,
\nonumber \\
{\Gamma'}_{12}^{(1)}&=&\ln\left(\frac{-t_1}{s}\right)
-\frac{1}{2}\ln\left(\frac{m^2}{s}\right)
+\frac{1}{2}L_{\beta}+\frac{\pi i}{2} \, , 
\nonumber \\
{\Gamma'}_{21}^{(1)}&=&\ln\left(\frac{-u_1}{s}\right)
-\frac{1}{2}\ln\left(\frac{m^2}{s}\right)
+\frac{1}{2}L_{\beta}+\frac{\pi i}{2} \, ,
\nonumber \\
{\Gamma'}_{22}^{(1)}&=&C_F\left[2\ln\left(\frac{-u_1}{s}\right)
-\ln\left(\frac{m^2}{s}\right)\right]-\frac{1}{2N_c}\left[2\ln
\left(\frac{-t_1}{s}\right)-\ln\left(\frac{m^2}{s}\right)
+L_{\beta}+\pi i\right] \, ,
\eeqa
where
$L_{\beta}=[(1-2m^2/s)/\beta] [\ln((1-\beta)/(1+\beta))+\pi i]$,
with $\beta=\sqrt{1-4m^2/s}$.
Note that all imaginary parts cancel out in Eq. (\ref{Auutt}).

\begin{figure}[htb] 
\epsfxsize=0.5\textwidth\epsffile{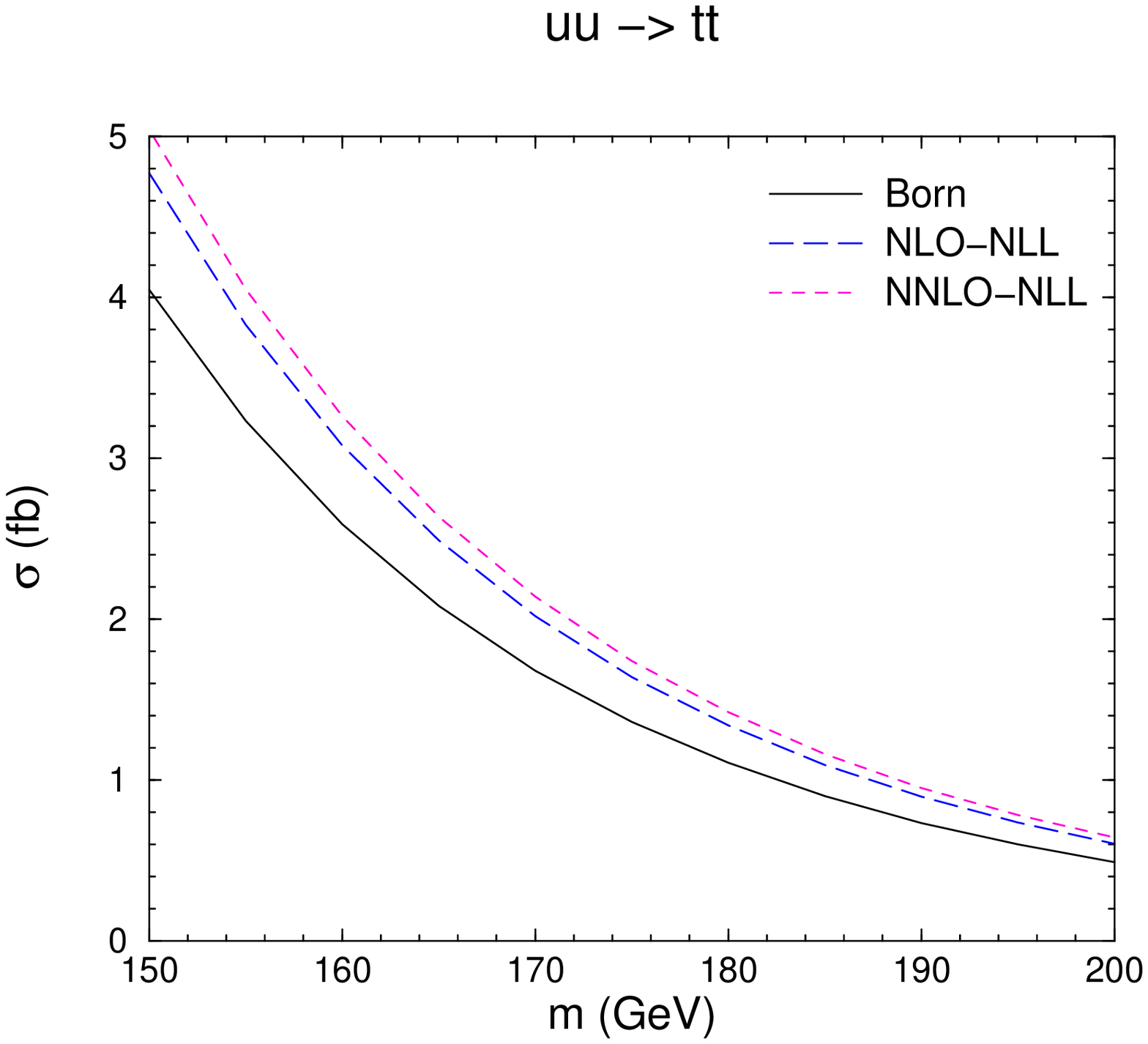}
\epsfxsize=0.5\textwidth\epsffile{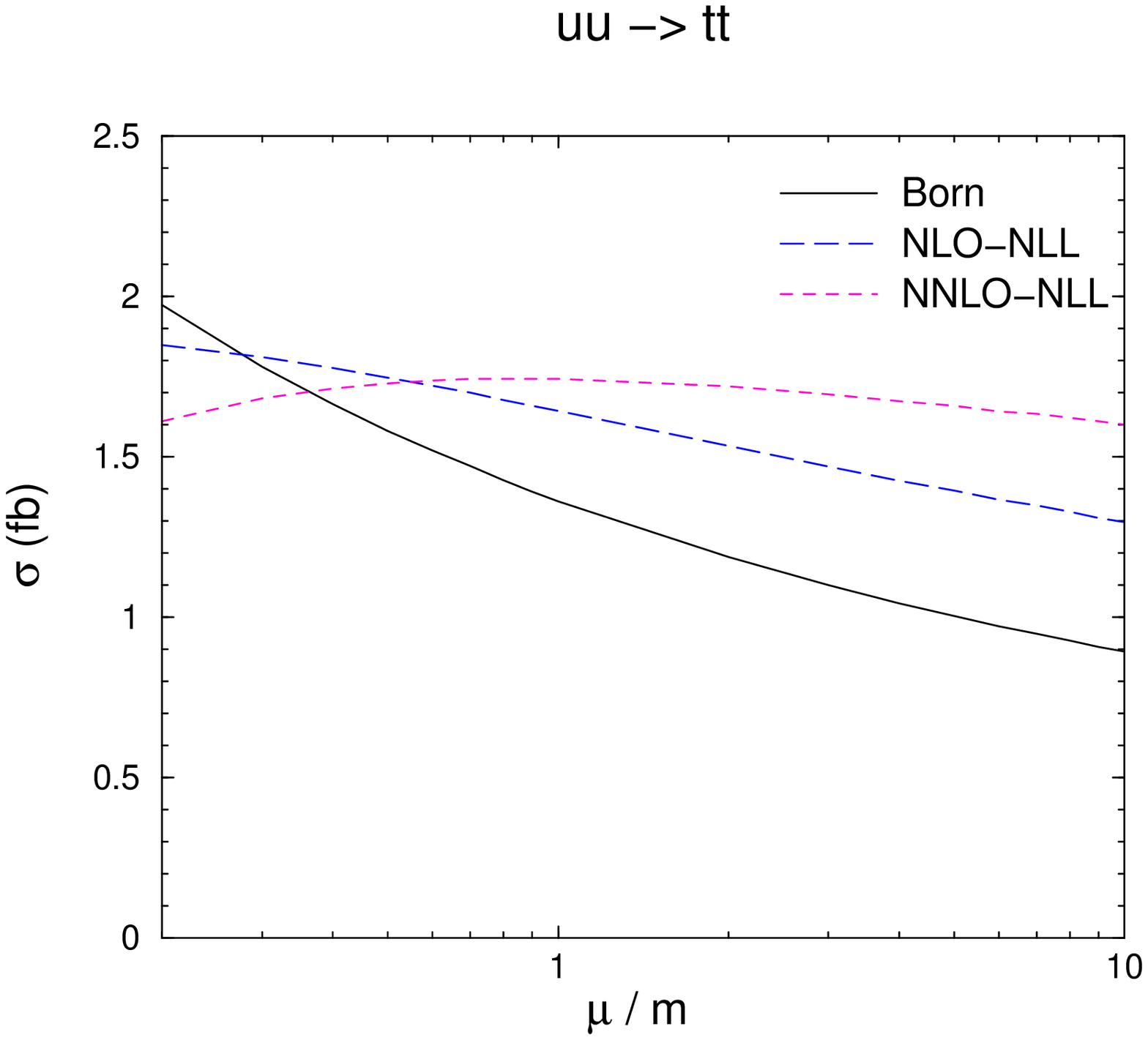}
\caption[]{The Born, NLO-NLL, and NNLO-NLL cross sections
for $uu\rightarrow tt$ in $p \bar p$ collisions
with $\sqrt{S}=1.96$ TeV and $\kappa_{\gamma}=\kappa_Z=0.1$.
Left:  cross section versus top-quark mass at scale $\mu=m$;
right: cross section versus  the   scale for $m= 175$~GeV.
}
\label{fcnc-cs-uutt} 
\end{figure}
The NNLO-NLL corrections for $uu \rightarrow tt$ are
\beqa
&& \frac{d^2{\hat\sigma}^{(2)}_{uu \rightarrow tt}}
{dt \, du}
=F^{uu \rightarrow tt}_B \frac{\alpha_s^2(\mu_R^2)}{\pi^2} 
\left\{\frac{1}{2} \left(c^{uu \rightarrow tt}_{3}\right)^2 
\left[\frac{\ln^3(s_4/m^2)}{s_4}\right]_+ \right.
\nonumber \\ && \hspace{-5mm}
{}+\left[\frac{3}{2} c^{uu \rightarrow tt}_{3} \, c^{uu \rightarrow tt}_{2}
-\frac{\beta_0}{4} c^{uu \rightarrow tt}_{3} \right] 
\left[\frac{\ln^2(s_4/m^2)}{s_4}\right]_+
\nonumber \\ && 
{}+\left[c^{uu \rightarrow tt}_{3} \, c^{uu \rightarrow tt}_{1}
+4C_F^2 \ln^2\left(\frac{\mu_F^2}{m^2}\right)
-4C_F T_2^{uu \rightarrow tt} \ln\left(\frac{\mu_F^2}{m^2}\right) \right.
\nonumber \\ && \quad \left.
{}+\frac{\beta_0}{4} c^{uu \rightarrow tt}_{3} 
\ln\left(\frac{\mu_R^2}{m^2}\right) 
-\zeta_2 \, \left(c^{uu \rightarrow tt}_{3}\right)^2 \right]
\left[\frac{\ln(s_4/m^2)}{s_4}\right]_+
\nonumber \\ &&  
{}+\left[-2C_F \ln\left(\frac{\mu_F^2}{m^2}\right) c^{uu \rightarrow tt}_{1}
-\frac{\beta_0}{2} C_F \ln\left(\frac{\mu_F^2}{m^2}\right)
\ln\left(\frac{\mu_R^2}{m^2}\right) \right.
\nonumber \\ && \quad \left. \left.
{}+C_F\frac{\beta_0}{4} \ln^2\left(\frac{\mu_F^2}{m^2}\right)
-\zeta_2 \, c^{uu \rightarrow t\gamma}_{2} \, c^{uu \rightarrow t\gamma}_{3}
+\zeta_3 \, \left(c^{uu \rightarrow t\gamma}_{3}\right)^2\right]
\left[\frac{1}{s_4}\right]_+ \right\} 
\nonumber \\ && 
+\frac{\alpha_s^2(\mu_R^2)}{\pi^2} \, A^{uu \rightarrow tt} \, 
\left\{\frac{3}{2} c^{uu \rightarrow tt}_3  
\left[\frac{\ln^2(s_4/m^2)}{s_4}\right]_+
-4C_F \ln\left(\frac{\mu_F^2}{m^2}\right) 
\left[\frac{\ln(s_4/m^2)}{s_4}\right]_+ \right.
\nonumber \\ && \quad \left.
-\zeta_2 \, c^{uu \rightarrow tt}_3 \, \left[\frac{1}{s_4}\right]_+
\right \} \, .
\label{NNLOuutt}
\eeqa

For the FCNC hadronic cross section $p(p_a)+{\bar p}(p_b) 
\rightarrow t(p_1)+t(p_2)$
we define $S=(p_a+p_b)^2$, $T=(p_a-p_1)^2$, and $U=(p_b-p_1)^2$,
and note that $p_{u_a}=x_a p_a$, $p_{u_b}=x_b p_b$.
The hadronic cross section is then given by
\beqa
\sigma^{FCNC}_{p{\bar p} \rightarrow tt}(S)&=&\int_{m^2-(S/2)
\left(1+\sqrt{1-4m^2/S}\right)}^{m^2-(S/2)\left(1-\sqrt{1-4m^2/S}\right)} dT 
\int_{-S-T+2m^2}^{m^2+m^2S/(T-m^2)} dU 
\nonumber \\ && \hspace{-28mm}
\times \int_{(m^2-T)/(S+U-m^2)}^1 dx_b \int_0^{x_b(S+U-m^2)+T-m^2} ds_4 \; 
\frac{x_a x_b}{x_b S+T-m^2} \, 
\phi(x_a) \, \phi(x_b) \, \frac{d^2{\hat\sigma}_{uu \rightarrow tt}}{dt \, du}
\eeqa
with
$x_a=\left[s_4-x_b(U-m^2)\right]/(x_b S+T-m^2)$.

\begin{figure}[htb] 
\centerline{\epsfxsize=0.5\textwidth\epsffile{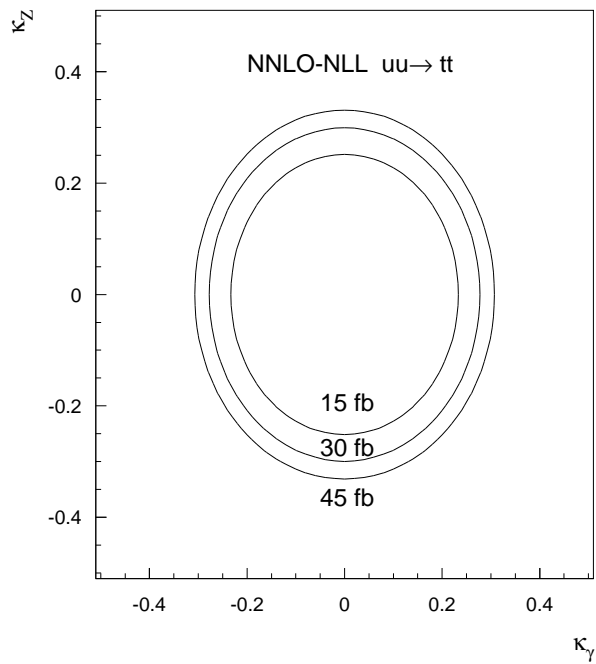}}
\caption[]{ Tevatron reach in $\kappa_Z - \kappa_\gamma$ 
           plane for the process $uu\to tt$ 
	   for cross section levels 15, 30, and 45 fb.}
\label{uutt-reach} 
\end{figure}

Our numerical results  are presented in  Fig.~\ref{fcnc-cs-uutt}.
The left frame presents  the cross section for the process 
$uu\rightarrow t t$ versus top-quark mass at the Tevatron, 
with $\sqrt{S}=1.96$ TeV, at Born, NLO-NLL, and NNLO-NLL
orders. One can see that the 
$\Delta\sigma_{NLO-NLL}$ and $\Delta\sigma_{NNLO-NLL}$ corrections are always
positive for this case for $\mu=m$.
The stabilisation of the NNLO-NLL cross section 
with respect to scale variation
is even more pronounced for this process compared to the two
previous cases as one can see in the
right frame of Fig.~\ref{fcnc-cs-uutt}. 
At $\mu=m=175$~GeV, we have for $\kappa_{\gamma}=\kappa_{Z}=0.1$:
$$\sigma_{Born}^{uu\rightarrow t t}          = 1.36 \ {\rm fb}, 
\ \Delta\sigma_{NLO-NLL}^{uu\rightarrow t t}= 0.28 \ {\rm fb}, 
\ \Delta\sigma_{NNLO-NLL }^{uu\rightarrow t t}= 0.10 \ {\rm fb}.$$
Taking into account the theoretical error due to the scale variation
between $m/2$ and $2m$ one has
$$\sigma_{NNLO-NLL }^{uu\rightarrow t t} =1.74^{+0.00}_{-0.02}\ {\rm fb}.$$

We can investigate the magnitude of the cross section if one of the
anomalous couplings is set to zero. If we use $\kappa_{\gamma}=0.1$
and set $\kappa_Z=0$ the NNLO-NLL cross section is 0.50 fb, while if we set
$\kappa_{\gamma}=0$ and use $\kappa_Z=0.1$ the NNLO-NLL cross section
is 0.38 fb.

We also note that we can have contributions from anomalous
couplings involving the charm quark.
The total cross section for the process $cc\rightarrow t t$ at the Tevatron,
using $\kappa_{tc\gamma}=\kappa_{tcZ}=0.1$, at $\mu=m=175$ GeV, is:
$\sigma_{NNLO-NLL }^{cc\rightarrow t t} 
=0.0044\ {\rm fb}$.
It is almost 3 orders of magnitude lower then the cross section for the
$uu\rightarrow t t$ process since both $c$-quarks originate from the sea.

The total cross section at the Tevatron for the mixed 
process $uc\rightarrow t t$,
using the value 0.1 for all anomalous couplings and $\mu=m=175$ GeV, is:
$\sigma_{NNLO-NLL }^{uc\rightarrow t t} 
=0.62\ {\rm fb}$.

Also, we note that ${\bar t} {\bar t}$ FCNC production at the Tevatron
has an equal cross section to $tt$ production, so if it is included
the total cross section is doubled.

Processes like same-sign top-quark production play a special role 
since $tt$ final states
give rise to same-sign leptons (SSL).
We have estimated that potential $uu\to b b W+ W+$ background 
can be safely neglected.
Assuming 10 fb$^{-1}$ integrated luminosity and $1\%$ 
efficiency (including $b-tagging$ and top-quark decay branchings)
one should require 30 fb of signal in order to observe 3 signal events
(to exclude signal at 95\% CL according to Poisson statistics
 under the assumption that the background is negligible).
We use this cross section as a rough idea on the order
of the signal cross section to which Tevatron will be sensitive.
Since the process $uu\rightarrow t t$ depends on $two$ parameters,
we present 15-30-45 fb contour levels of  the $uu\rightarrow t t$ 
cross section in  $\kappa_{\gamma}-\kappa_{Z}$ plane 
in Fig.~\ref{uutt-reach}. This figure, which gives a rough idea about 
the Tevatron reach, shows that the Tevatron sensitivity is just slightly better 
for the $\kappa_{\gamma}$ coupling.

At this point, it is worth mentioning the present experimental limits
on the $\kappa_{\gamma}$, $\kappa_{Z}$ couplings.
First, a limit on FCNC top-quark couplings has been established
by the CDF collaboration in terms of the limit on the branching ratio of 
FCNC top-quark decay~\cite{Abe:1997fz}: 
$Br(t\to q\gamma)<3.2\%$, $Br(t\to qZ)<33\%$
which corresponds to  $\kappa_{\gamma}<0.14$, $\kappa_{Z}<0.69$
limit in the notation of Eq.~(\ref{fcnc-eq}).
A more stringent limit on  $\kappa_{\gamma}<0.10$
has been established by the L3 collaboration~\cite{Achard:2002vv}.
Finally, the best limits on $\kappa_{\gamma}$ have been established by 
H1 ($\kappa_{\gamma}<0.090$)~\cite{HERA4,HERA5} and 
by ZEUS ($\kappa_{\gamma}<0.058$)~\cite{HERA1,HERA2,HERA3} collaborations.
All the limits quoted above correspond to the notation of Eq.~(\ref{fcnc-eq}).
One can see that in light of the present experimental limits, the 
$uu\to tt$ process is unlikely to be observed  at the Tevatron
unless $uu\to tt$ production involves chromomagnetic flavor-changing current,
which is beyond the scope of the present paper.
In order to estimate Tevatron sensitivity to 
$p\bar{p}\to tZ$, $p\bar{p}\to t\gamma$ processes one should study in details
all potential backgrounds which is, again, beyond the scope of our paper.

\mysection{$eu \rightarrow et$}

The last process we present is single-top production at HERA 
\cite{HERA1,HERA2,HERA3,HERA4,HERA5}.
We have previously studied this process in Ref. \cite{fcnc-hera}
where we calculated the NLO soft corrections with $tq\gamma$
couplings. Here we extend our previous calculation
by also including  $tqZ$ couplings and NNLO-NLL soft corrections. 

\begin{figure}[htb] 
\centerline{\epsfxsize=0.6\textwidth\epsffile{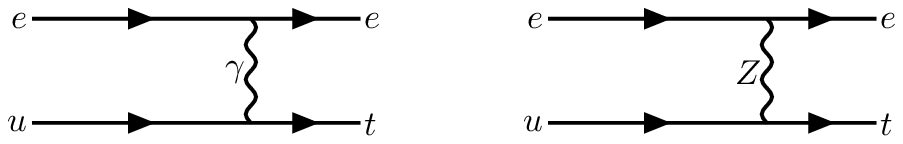}}
\caption[]{Tree-level Feynman diagrams for the process 
$eu \rightarrow e t$.}
\label{fig-et} 
\end{figure}

For the process $e(p_e)+u(p_u) \rightarrow e(p_f)+t(p_t)$,
we define the kinematical invariants $s=(p_e+p_u)^2$,
$t=(p_t-p_u)^2$, $u=(p_t-p_e)^2$, and $s_4=s+t+u-m^2-2m_e^2$,
with $m$ the top-quark mass and $m_e$ the electron mass.
Note that near threshold $s_4 \rightarrow 0$.
The respective tree-level Feynman diagrams are shown in Fig.~\ref{fig-et}.

The differential Born cross section is 
\beq
\frac{d^2{\hat\sigma}^{eu \rightarrow et}_B}{dt \; du}
=F^{eu \rightarrow et}_B \delta(s_4)
\eeq
where 
\beqa
F^B_{eu\rightarrow et}&=& \frac{\pi \, \alpha^2}
{m^2 \, (s-m_e^2)^2}
\left[ 
\left\{-t\left[2m_e^4+m^4-2s^2+(2s+t)(2s-m^2-2m_e^2)\right]
-2m_e^2m^4\right\} 
\nonumber\right. \\ 
&& \times
\left\{\frac{8 \kappa_{\gamma}^2}{t^2}
+\frac{4 \kappa_{\gamma}\kappa_Z(1 - 4 \sin^2\theta_W)}
{\sin\theta_W \, \cos\theta_W \, t \, (t-m_Z^2)}
+\frac{\kappa_Z^2(1 - 4 \sin^2\theta_W+8\sin^4\theta_W)}
{\sin^2\theta_W \, \cos^2\theta_W \, (t-m_Z^2)^2}\right\}
\nonumber
\\
&&
\left.
-\frac{\kappa_Z^2 m_e^2 ( -2 m^4 + t  m^2 + t^2)}{\sin^2\theta_W \, \cos^2\theta_W \, (t-m_Z^2)^2}
\right]
\, ,
\eeqa
with $\theta_W$ the weak-mixing angle,
$\kappa_{\gamma} \equiv \kappa_{tu\gamma}$, and
$\kappa_{Z} \equiv \kappa_{tuZ}$. 

The NLO-NLL corrections for $e u \rightarrow et $ are
\beq
\frac{d^2{\hat\sigma}^{(1)}_{eu\rightarrow et}}{dt \, du}
=F^{eu \rightarrow et}_B 
\frac{\alpha_s(\mu_R^2)}{\pi} \left\{
c^{eu \rightarrow et}_{3} \left[\frac{\ln(s_4/m^2)}{s_4}\right]_+
+c^{eu \rightarrow et}_{2} \left[\frac{1}{s_4}\right]_+
+c^{eu \rightarrow et}_{1}  \delta(s_4)\right\} \, .
\label{NLOeuet}
\eeq
Here $c^{eu \rightarrow et}_{3}=2C_F$,
\beqa
c^{eu \rightarrow et}_{2}&=& 2{\Gamma'}_S^{(1)}
-C_F\left[1+2\ln\left(\frac{-u+m_e^2}{m^2}\right)
+\ln\left(\frac{\mu_F^2}{s}\right)\right]
\nonumber \\
& \equiv & T^{eu \rightarrow et}_{2}-C_F
\ln\left(\frac{\mu_F^2}{m^2}\right) \, , 
\eeqa
where
${\Gamma'}_S^{(1)}=C_F \ln[(m^2-t)/(\sqrt{s} m)]$ is the one-loop
soft anomalous dimension and 
$T^{eu \rightarrow et}_{2}$ is the scale-independent part of
$c^{eu \rightarrow et}_{2}$, 
and
\beq
c^{eu \rightarrow et}_{1}=
\left[-\frac{3}{4} 
+\ln\left(\frac{-u+m_e^2}{m^2}\right)\right]
C_F\ln\left(\frac{\mu_F^2}{m^2}\right) 
\eeq
is the scale-dependent part of the $\delta(s_4)$ terms.

The NNLO-NLL corrections for $eu \rightarrow et$ are
\beqa
&& \hspace{-5mm}\frac{d^2{\hat\sigma}^{(2)}_{eu \rightarrow et}}
{dt \, du}
=F^{eu \rightarrow et}_B \frac{\alpha_s^2(\mu_R^2)}{\pi^2} 
\left\{\frac{1}{2} \left(c^{eu \rightarrow et}_{3}\right)^2 
\left[\frac{\ln^3(s_4/m^2)}{s_4}\right]_+ \right.
\nonumber \\ && \hspace{-5mm}
{}+\left[\frac{3}{2} c^{eu \rightarrow et}_{3} \, c^{eu 
\rightarrow et}_{2}
-\frac{\beta_0}{4} c^{eu \rightarrow et}_{3} \right] 
\left[\frac{\ln^2(s_4/m^2)}{s_4}\right]_+
\nonumber \\ && \hspace{-5mm}
{}+\left[c^{eu \rightarrow et}_{3} \, c^{eu \rightarrow et}_{1}
+C_F^2\ln^2\left(\frac{\mu_F^2}{m^2}\right)
-2C_F T_2^{eu \rightarrow et}\ln\left(\frac{\mu_F^2}{m^2}\right)
\right.
\nonumber \\ && \quad \left.
{}+\frac{\beta_0}{4} c^{eu \rightarrow et}_{3} 
\ln\left(\frac{\mu_R^2}{m^2}\right)
-\zeta_2 \, \left(c^{eu \rightarrow et}_{3}\right)^2 \right]
\left[\frac{\ln(s_4/m^2)}{s_4}\right]_+
\nonumber \\ && \hspace{-5mm} 
{}+\left[-C_F \ln\left(\frac{\mu_F^2}{m^2}\right)
c^{eu \rightarrow et}_{1}
-\frac{\beta_0}{4} C_F \ln\left(\frac{\mu_F^2}{m^2}\right) 
\ln\left(\frac{\mu_R^2}{m^2}\right) \right.
\nonumber \\ && \quad \left. \left.
{}+C_F \frac{\beta_0}{8} \ln^2\left(\frac{\mu_F^2}{m^2}\right)
-\zeta_2 \, c^{eu \rightarrow et}_{2} \, c^{eu \rightarrow et}_{3}
+\zeta_3 \, \left(c^{eu \rightarrow et}_{3}\right)^2\right]
\left[\frac{1}{s_4}\right]_+ \right\} \, .
\label{NNLOeuet}
\eeqa

For the FCNC hadronic cross section $e(p_e)+p(p_p) 
\rightarrow e(p_f)+t(p_t)$
we define $S=(p_e+p_p)^2$, $T=(p_t-p_p)^2$, and $U=(p_t-p_e)^2$,
and note that $p_u=x p_p$.
The hadronic cross section is then given by
\beqa
\sigma^{FCNC}_{ep \rightarrow et}(S)&=&\int_{T_{min}}^{T_{max}} dT
\int_{-S-T+m^2+2m_e^2}^{U_{max}} dU 
\int_0^{S+T+U-m^2-2m_e^2} ds_4
\nonumber \\ &&
\times \frac{x}{S+T-m^2-m_e^2} \;
\phi(x) \, \frac{d^2{\hat\sigma}_{eu \rightarrow et}}{dt \, du}
\eeqa
with 
\beq
T_{^{max}_{min}}=m^2-\frac{(S-m_e^2)}{2S}
\left[S+m^2-m_e^2 \mp \sqrt{\left(S-m^2-m_e^2\right)^2-4m^2m_e^2}\right]\, ,
\eeq
$U_{max}=2m^2+m_e^2-T-S(m^2-T)/(S-m_e^2)-(S-m_e^2)m^2/(m^2-T)$,
and $x=(s_4+m_e^2-U)/(S+T-m^2-m_e^2)$.

\begin{figure}[htb] 
\epsfxsize=0.5\textwidth\epsffile{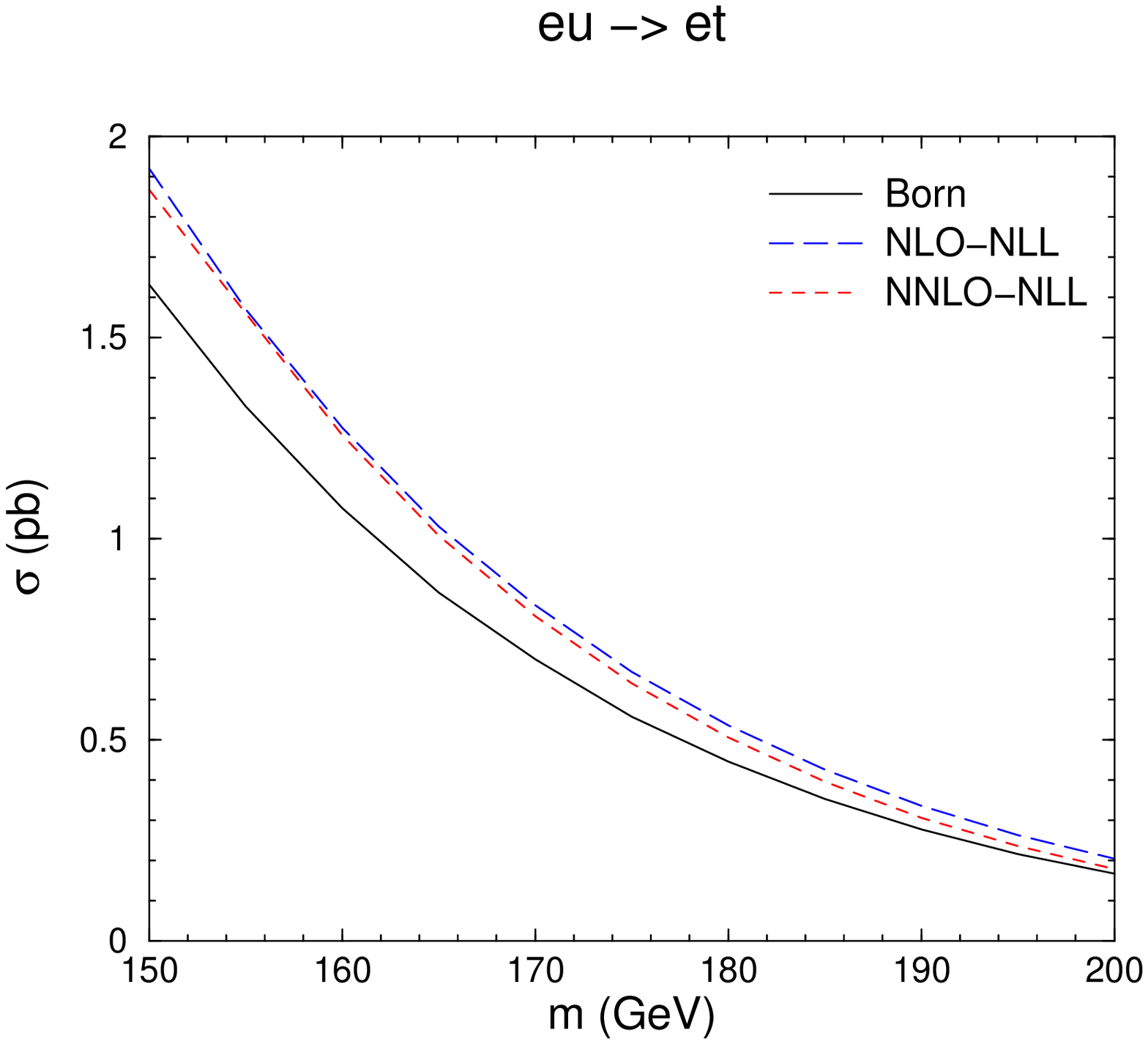}
\epsfxsize=0.5\textwidth\epsffile{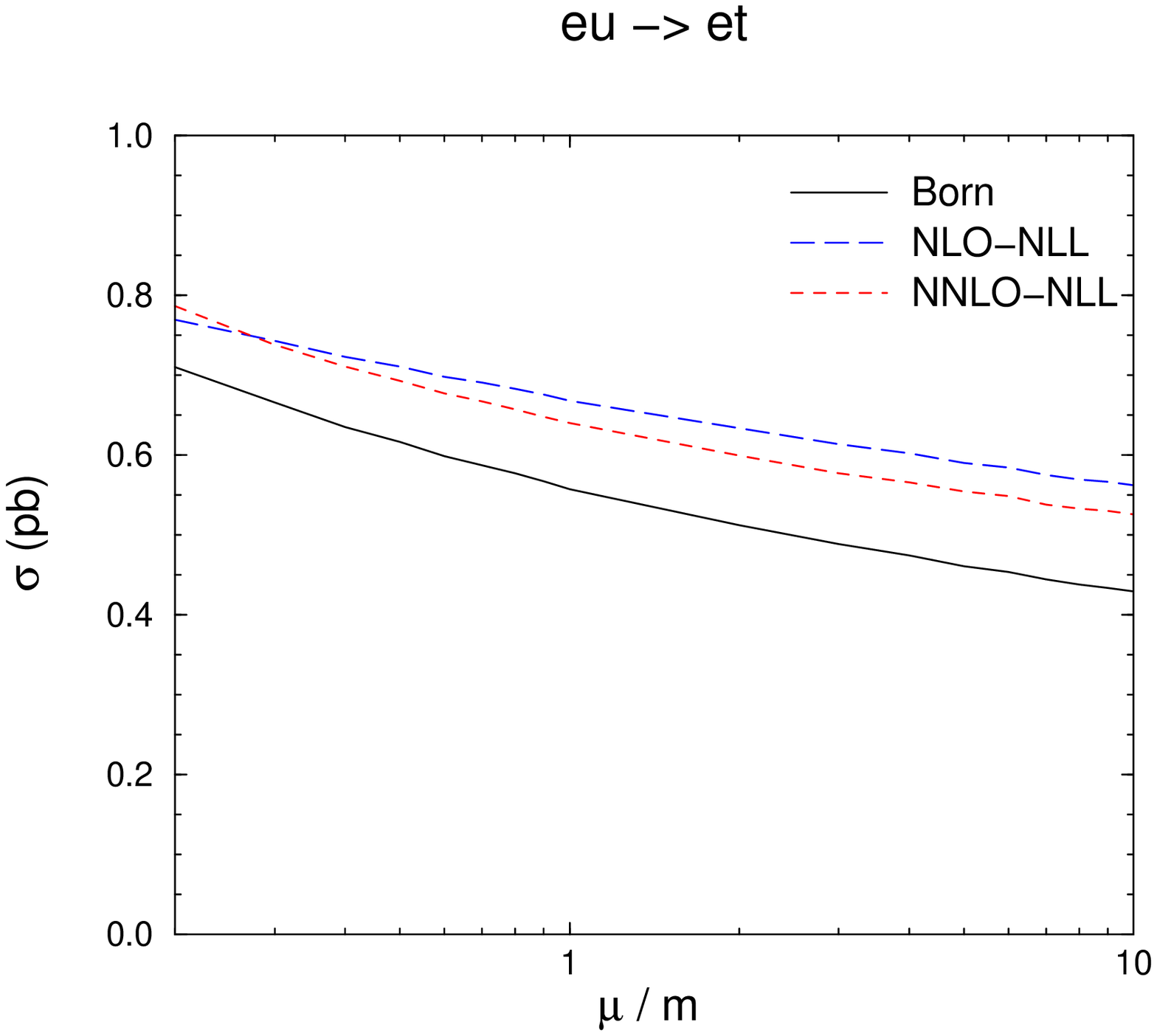}
\caption[]{The Born, NLO-NLL, and NNLO-NLL cross sections 
for $eu\rightarrow et$ at HERA with $\sqrt{S}=318$ GeV
and $\kappa_{\gamma}=\kappa_Z=0.1$.
Left:  cross section versus top-quark mass at scale $\mu=m$;
right: cross section versus  the   scale for $m= 175$~GeV.
}
\label{figeuet} 
\end{figure}

Our numerical results  are presented in  Fig.~\ref{figeuet}.
The left frame presents  the cross section for the process 
$eu\rightarrow et$ versus top-quark mass at HERA, 
with $\sqrt{S}=318$ GeV, at Born, NLO-NLL, and NNLO-NLL
orders. One can see that the 
$\Delta\sigma_{NLO-NLL}$ corrections are positive while 
the  $\Delta\sigma_{NNLO-NLL}$ corrections are negative
for this case for $\mu=m$.
The stabilisation of the NNLO-NLL cross section with respect to
scale variation is shown in the right frame of Fig.~\ref{figeuet}. 
At $\mu=m=175$~GeV, we have for $\kappa_{\gamma}=\kappa_{Z}=0.1$:
$$\sigma_{Born}^{eu\rightarrow et}          = 0.56 \ {\rm pb}, 
\ \Delta\sigma_{NLO-NLL}^{eu\rightarrow et}=  0.11\ {\rm pb}, 
\ \Delta\sigma_{NNLO-NLL }^{eu\rightarrow et}= -0.03 \ {\rm pb}.$$
Taking into account the theoretical error due to the scale variation
between $m/2$ and $2m$ one has
$$\sigma_{NNLO-NLL }^{eu\rightarrow et} =0.64^{+0.05}_{-0.04}\ {\rm pb}.$$

We note that almost all of the contribution comes from
the $\kappa_{\gamma}$ coupling. If we use $\kappa_{\gamma}=0.1$
and set $\kappa_Z=0$ the cross section 
remains practically unchanged (i.e. 0.64 pb at NNLO-NLL), while if we set
$\kappa_{\gamma}=0$ and use $\kappa_Z=0.1$ the NNLO-NLL cross section
is only 0.0018 pb.

\begin{figure}[htb]
\begin{center} 
\epsfxsize=0.5\textwidth\epsffile{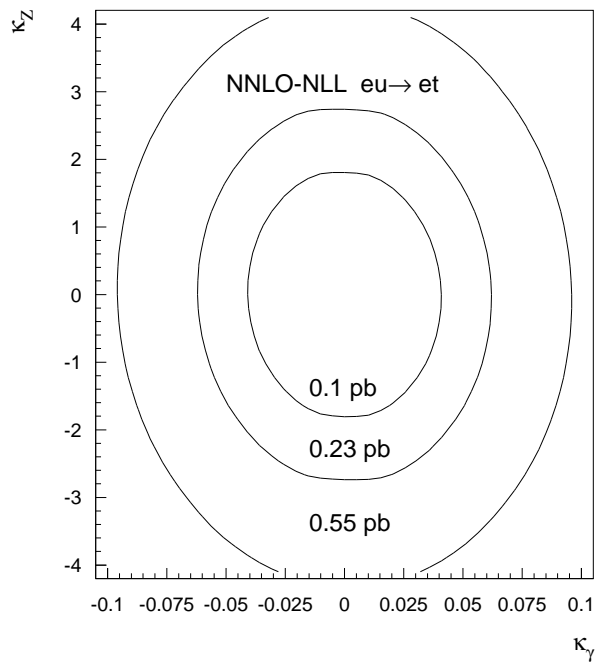}
\vspace*{-0.5cm}
\caption{\label{uete-reach} 
           HERA reach in $\kappa_Z - \kappa_\gamma$ 
           plane for the process $eu\to et$ 
	   for cross section levels 0.55, 0.23, and 0.1 pb.}
\end{center} 
\end{figure}

In Fig.~\ref{uete-reach} we present
contour levels for the $eu\to et$ process
in $\kappa_\gamma-\kappa_Z$ plane.
One of the contours is at 0.55 pb level
which is  the present  HERA limit
reported by H1~\cite{HERA4,HERA5} collaboration.
The second contour is for 0.23 pb level
which corresponds to the limit
reported by ZEUS~\cite{HERA3} 
collaboration.
The third  contour is for
0.1 pb level anticipating the improvement
of the experimental limits for higher luminosities of
the coming HERA II stage.  Fig.~\ref{uete-reach}
clearly demonstrates 
that the HERA collider (in contrast to the $uu\to tt$
process at the Tevatron)
is much more (about a factor of 40) sensitive to the 
$\kappa_{\gamma}$ coupling than to the $\kappa_Z$ one.

We also note that we can have contributions from anomalous
couplings involving the charm quark. Using
$\kappa_{tc\gamma}=\kappa_{tcZ}=0.1$ the cross section at HERA for
$ec \rightarrow et$, with $\mu=m=175$ GeV, is 
$\sigma_{NNLO-NLL }^{ec\rightarrow et}=0.0020$ pb, 
which is low due to the small sea charm-quark densities.

In the case of $e\bar t$ production, involving the anti-top, 
the cross section is small,
$\sigma_{NNLO-NLL }^{e{\bar u}\rightarrow e{\bar t}}=0.0079$ pb
at $\mu=m=175$ GeV, 
and thus asymmetrical to $et$ production, in contrast to the other
processes which we considered at the Tevatron. This is because
at HERA the valence up-quark density can not contribute 
to $e\bar t$ production.

\mysection{Conclusions}

We have studied several FCNC processes at the Tevatron and HERA
colliders involving top quark production via anomalous $tqV$ couplings,
with $q$ an up or charm quark and $V$ a photon or $Z$-boson. 
The partonic processes involved are $gu \rightarrow tZ$,
$gu \rightarrow t \gamma$, $uu \rightarrow tt$, and 
$eu \rightarrow et$ 
as well as the corresponding ones with the
up quark replaced by a charm quark. We have calculated NLO and 
NNLO soft-gluon corrections to these cross sections.

The Tevatron and HERA colliders provide a unique opportunity
for precise study of  FCNC couplings involving top-quark production.
For some of these processes the uncertainty of the Born-level cross sections
due to the choice of the QCD scale could be of the order of
 100\%.
The search for FCNC single top-quark production at HERA is especially
intriguing since a very recent experimental study of this process not
only establishes a limit on the anomalous FCNC coupling  but also
presents an excess of events in the leptonic channel over the SM
background (5 events observed while $1.31 \pm 0.22$ events are expected
for SM background !)~\cite{HERA5}.

It is also worth mentioning that Tevatron and HERA are highly complementary
in constraining FCNC top-quark couplings:
HERA has the best potential in limiting the $\kappa_\gamma$ coupling
and  is practically insensitive to  $\kappa_Z$,
while the Tevatron is sensitive to both  $\kappa_Z$ and $\kappa_\gamma$ 
couplings. Therefore  the combination of $\kappa_\gamma$ limit from HERA
and $\kappa_Z$ limit from the Tevatron will be  the most stringent limit
in $\kappa_\gamma$ - $\kappa_Z$ plane.

The main outcome of this paper is that
we have found that higher-order QCD  corrections stabilize
the factorization and renormalization scale dependence of the 
FCNC cross sections down to the level of about 10\% or less,
and provide modest to significant enhancements to the leading
order result.

\section*{Acknowledgements}
We would like to thank Jeff Owens for useful discussions.
The research of N.K. has been supported by a Marie Curie Fellowship of 
the European Community programme ``Improving Human Research Potential'' 
under contract number HPMF-CT-2001-01221.
The work of A.B. was supported in part by the U.S. Department of Energy.

\end{document}